# Distance function wavelets – Part III: "Exotic" transforms and series

W. Chen

Simula Research Laboratory, P. O. Box. 134, 1325 Lysaker, Norway

E-mail: wenc@simula.no

(8 June 2002)

**Summary**

Part III of the reports consists of various unconventional distance function wavelets (DFW). The dimension and the order of partial differential equation (PDE) are first used as a substitute of the scale parameter in the DFW transform and series, especially with the space and time-space potential problems. It is noted that the recursive multiple reciprocity formulation is the DFW series. The Green second identity is used to avoid the singularity of the zero-order fundamental solution in creating the DFW series. The fundamental solutions of various composite PDEs are found very flexible and efficient to handle a broad range of problems. We also discuss the underlying connections between the crucial concepts of dimension, scale and the order of PDE through the analysis of dissipative acoustic wave propagation. The shape parameter of the potential problems is also employed as the "scale parameter" to create the non-orthogonal DFW. This paper also briefly discusses and conjectures the DFW correspondences of a variety of coordinate variable transforms and series. Practically important, the anisotropic and inhomogeneous DFW's are developed by using the geodesic distance variable. The DFW and the related basis functions are also used in making the kernel distance sigmoidal functions, which are potentially useful in the artificial neural network and machine learning. As or even worse than the preceding two reports, this study scarifies mathematical rigor and in turn unfetter imagination. Most results are intuitively obtained without rigorous analysis. Follow-up research is still under way. The paper is intended to inspire more research into this promising area.

## 1. Introduction

The report is the third in series [1,2] about my latest advances on the distance function wavelets (DFW). One of the main motivations behind this research is to extend the DFW to various scale-invariant potential problems where unlike the Helmholtz and convection-diffusion problems, the standard scale parameter does not exist. In section 2, we instead use the dimension as a substitute of the scale parameter to develop the space and time-space potential DFW's, whose transforms involve continuous dimension variation (fractal and multifractal). The links between these DFW's and the Kontorovich-Lebedev and Mellin transforms are pointed out. In section 3, it is noted that the boundary particle method (BPM) [3,4] and recursive multiple reciprocity BEM have the DFW series formulation, where the "scale parameter" is interpreted as the order of the high-order general solutions and fundamental solutions of partial differential equations (PDE). Its corresponding DFW transforms involve the fractional derivative and complex-order derivative. The singularity of the zero-order fundamental solution is a troublesome specter in creating the DFW series in terms of the order of PDE. Section 4 applies the Green second identity to circumvent this thorny problem. It is also noted that the recursive multiple reciprocity method can become very efficient via the scale orthogonality. The fundamental solutions of various composite PDEs are also presented to handle a wide range of problems. The technique is called the composite multiple reciprocity. However, in some cases like the convection-diffusion equation, the reciprocity principle does not hold. Thus, the differentiation smoothing is a better term for this approach. The time-space DFW series is also proposed. It is discussed that the composite multiple reciprocity DFW could lead to a boundary-only DFW technique not only for numerical PDE but also for general data processing. In section 5, the underlying connections between the essential concepts of dimension, scale and the order of PDE are critically discussed through the analysis of the frequency-dependent attenuation of acoustic wave propagation. Furthermore, section 6 employs the shape parameter of various rotational and translate invariant MQ-type (the MQ is the abbreviation of the multiquadratic) distance functions as the "scale parameter" to construct the pre-wavelets. Section 7 mirrors the DFW on a variety of existing standard integral transforms and series. In section 8, we introduce practically important geodesic DFW, which uses the geodesic distance variable and will be potentially very useful in handling the anisotropic and inhomogeneous problems. Section 9 employs the DFW and the related basis functions to make the kernel distance sigmoidal functions for the artificial neural network and machine learning. Finally, section 10 gives a brief remark on the DFW promise. "Exotic" in the title means that the DFW's introduced in this report use the unconventional dimension, the order of PDE, shape parameter as a substitute of the scale parameter. As or even worse than the preceding reports [1,2], this research lacks mathematical rigor and is contemplated to be future-orientated in the best hope leading to an enormous territory rich in open problems.

## 2. Potential DFW transform using dimension parameter

Unlike the Helmholtz and convection-diffusion problems, the kernel solutions of various potential PDEs, e.g. the Laplacian, are scale invariant and have not explicit scale provision within them. Notwithstanding, in terms of dimensional parameter as a substitute of the scale parameter, this section aims to create the potential DFW via the PDE kernel solutions of space and time-space potential problems, followed by section 3 to present the multiple reciprocity DFW using the order of PDE as the scale argument, and then, section 4 derives the multiple reciprocity (MR) DFW series. Section 5 provides a unified mathematical and physical view on dimensionality, PDE order and scale parameters and their effects. The DFW using the dimension argument may find some applications in high-dimensional problems such as the neural network.

**2.1. Space potential DFW transform**

Without a loss of generality, let us consider the fundamental solution of the Laplace equation [5]

$$u_{L_n}^*(r_k) = \begin{cases} -\dfrac{1}{S_1} r_k, & n = 1 \\ -\dfrac{1}{S_2} \ln r_k, & n = 2 \\ -\dfrac{r_k^{2-n}}{(n-2)S_n}, & n \geq 3 \end{cases} \quad (1)$$

where $r_k = \|x - x_k\|$, $S_n$ is the surface area of the unit sphere given by

$$S_1 = 2, \qquad S_2 = 2\pi, \qquad S_n = \frac{2\pi^{n/2}}{\Gamma(n/2)}. \quad (2)$$

$\Gamma$ is the gamma function. We can verify (1) via the radial Laplacian

$$\frac{d^2 u_{L_n}^*}{dr^2} + \frac{n-2}{r} \frac{du_{L_n}^*}{dr} = -\Delta_i. \quad (3)$$

In terms of the fractal geometry and fractal Hausdorff dimension, $n$ is not necessarily an integer. We note that

$$u_{L_n}^*(r_k) = \begin{cases} -\dfrac{1}{S_2} \ln r_k, & n = 2 \\ -\dfrac{r_k^{2-n}}{(n-2)S_n}, & n \neq 2 \end{cases} \quad (4)$$

could satisfy equation (3) with arbitrary real-valued *n*. It is the author's guess that two dimensions are special since the Laplacian happens to be the second order of partial differential equation model. A fraction *n* may suggest the fractal field and potential or the fractional Laplace operator [6,7].

The Laplacian is scale invariance, which happens to agree with the definition of the self-similarity and fractal dimension. In order to create the continuous DFW transform, it is necessary to use the concept of continuous dimension variation instead of conventional integer quantum jumps of dimensions. In terms of Laplacian fundamental solution (4), we have the continuous DFW:

$$P(n,\xi) = \int_{IR^n} f(x) \frac{u_{L_n}^*(\|\xi - x\|)^{-1}}{\|\xi - x\|} dx \qquad (5a)$$

and

$$f(x) = \frac{1}{C_P} \int_0^{+\infty} \int_{IR^n} P(n,\xi) u_{L_n}^*(\|x - \xi\|) n d\xi dn . \qquad (5b)$$

where $C_p$ can be determined by the continuous wavelets theory. It is known that negative dimension occurs in the exploration of some hidden physics secrets. It is therefore reasonable to extend dimension *n* to negative real number space. Furthermore, the complex space has been presented in literature. That means *n* could even be a complex number. It is also noted that the derivative of the Laplacian fundamental solutions of any dimensions with respect to the distance has the same form

$$\frac{du_{L_n}^*(r_k)}{dr_k} = \frac{1}{S_n r_k^{n-1}} . \qquad (6)$$

(6) can be used to replace the Laplacian fundamental solution in (5) to construct the double layer potential DFW.

In the Helmholtz-Laplace transforms [1,2], we use the fundamental and general solutions of the modified Helmholtz equation

$$w_1(\mu r_k) = \frac{\mu^{1/2}}{2} e^{-\mu r_k}, \qquad (7a)$$

$$w_n(\mu r_k) = \frac{\mu^{n-1/2}}{2\pi}(2\pi\mu r_k)^{-(n/2)+1} K_{(n/2)-1}(\mu r_k), \quad n \neq 1; \qquad (7b)$$

$$\hat{w}_1(\mu r_k) = \frac{\mu^{1/2}}{2} e^{\mu r_k}, \qquad (8a)$$

$$\hat{w}_n(\mu r_k) = \frac{\mu^{n-1/2}}{2\pi}(2\pi\mu r_k)^{-(n/2)+1} I_{(n/2)-1}(\mu r_k), \quad n \neq 1, \qquad (8b)$$

where $I$ and $K$ denote the modified Bessel function of the first and second kinds. Instead of using the parameter $\mu$, we use the dimension parameter $n$ as the "scale parameter" and have the DFW transform for a suitable function $f(x)$

$$H(n,\xi) = \int_{IR^n} f(x) \frac{w_n(\mu\|\xi - x\|)}{\|\xi - x\|} dx, \qquad (9a)$$

$$f(x) = \frac{1}{C_w} \int_{\gamma-i\infty}^{\gamma+i\infty} \int_{IR^n} H(n,\xi) \hat{w}_n(\mu\|x - \xi\|) n\, d\xi\, dn. \qquad (9b)$$

Note that the dimensionality here is the complex number. In contrast, the classic forward and inverse Kontorovich-Lebedev transforms [8] are defined by

$$L(\mu,\alpha) = \int_0^\infty K_\alpha(\mu x) f(x) \frac{dx}{x}, \qquad (10a)$$

$$f(x) = \frac{1}{i\pi} \int_{-i\infty}^{i\infty} \alpha I_\alpha(\mu x) L(\mu, \alpha) d\alpha, \qquad (10b)$$

where the imaginary index $\alpha$ of $K_\alpha$ matches the dimensionality in the DFW transform (9). As $\mu \to 0$, the modified Helmholtz equation degenerates into the Laplace equation and the Kontorovich-Lebedev transforms become the Mellin transforms, i.e.

$$M(\alpha) = \int_0^\infty x^{\alpha-1} f(x) dx, \qquad (11a)$$

$$f(x) = \frac{1}{2\pi i} \int_{c-i\infty}^{c+i\infty} x^{-\alpha} M(\alpha) d\alpha. \qquad (11b)$$

It is now clear that the Mellin transform has the lurking background of the Laplace equation. In terms of the complex dimensionality, the inverse transform (5b) is alternatively replaced by

$$f(x) = \frac{1}{\hat{C}_P} \int_{c+i\infty}^{c+i\infty} \int_{IR^n} P(n, \xi) u_{L_n}^*(\|x - \xi\|) n d\xi dn. \qquad (12)$$

The Laplacian potential DFW transform (5a) and (12) may be considered the DFW counterpart of the Mellin transform which finds applications in scale-invariant image and speech recognition. By analogy with the Mellin-Fourier series [9], we can construct the expansion series of the Laplacian DFW. It is the author's guess that the existence condition for the Laplacian DFW is similar to that of the Mellin transform, i.e.

$$\int_{IR^n} |f(x)| u_{L_n}^*(\|\xi - x\|) dx \prec \infty. \qquad (13)$$

We may consider the Kontorovich-Lebedev and Mellin transforms are the degenerate DFW transforms respectively using the solutions of the modified Helmholtz and Laplace equations as the Hankle transform is to the Helmholtz-Fourier transform. In the later section 3, we will show that the solutions of high-order Laplacian can also be used to create the DFW transforms of this kind.

## 2.2. Time-space potential DFW transform

The time-space potential can be seen as an extension of the spatial potential under the general Euclidean-like space. The origin of the universe, as stated by the big bang theory, starts from a singularity (zero dimension) to time dimension to one spatial dimension to two and three spatial dimensions. If we speculate this cosmos dimension evolution is a continuous (or finite quantum jumps but integrable) process, the continuous time-space DFW could come into play. Unfortunately, as far as the author knows, the temporal fractal fundamental solution of potential PDE is not available albeit in principle existing and solvable. So, the time-space potential DFW transforms developed below are in fact the space potential DFW augmented with one time dimension. It is stressed that besides the finite four time-space dimensions, the dimensionality could reach much higher and even infinite in terms of other physical parameters. The dimensional DFW also has close ties with the multifractal concept.

For time-dependent problems, we typically meet the so-called "heat potential" and "wave potential" [5]. Consider the diffusion equation

$$\frac{\partial u}{\partial t} - \kappa \nabla^2 u = f(x,t), \tag{14}$$

where $\kappa$ is the conductivity parameter of media in terms of heat problems, the heat potential is defined by

$$u(x,t) = \int_0^t \int_{IR^n} u_{h_n}^*(x-\xi, t-\tau) f(\xi,\tau) d\xi d\tau, \tag{15}$$

where $f$ is a distribution that vanish for $\tau<0$, and

$$u_{h_n}^*(x-\xi, t-\tau) = \frac{H(t-\tau)}{[4\pi\kappa(t-\tau)]^{n/2}} e^{-\|x-\xi\|^2/4\kappa(t-\tau)}, \tag{16}$$

where $H$ is the Heaviside step function, which is added to ensure the causality. In terms of the heat potential (15), we can make the time-space DFW transforms

$$H(n,\xi,\tau) = \int_0^{+\infty} \int_{IR^n} f(x,t) \frac{u_{h_n}^*(\xi-x,\tau-t)^{-1}}{\|\xi-x\|} dx dt \qquad (17a)$$

and

$$f(x,t) = \frac{1}{C_H} \int_{c-i\infty}^{c+i\infty} \int_0^t \int_{IR^n} H(n,\xi,\tau) u_{h_n}^*(x-\xi,t-\tau) n d\xi d\tau dn. \qquad (17b)$$

It is known that the Black-Scholes PDE model for financial option is also a diffusion-type equation usually involving multiple variables (up to thousands dimensions). The Monte Carlo method is now the dominant simulation technique in this field. The dimensional DFW may play a role for analyzing this high dimension problem. Through the dimensional DFW transforms, we may find which dimensions in very high-dimensional problems play the key role in particular objects so that a "dimension compression" may be achieved.

If $\kappa$ in (14) is purely imaginary such that $\kappa = \dfrac{i\hbar}{2m}$, where $m$ is the mass of the quantum particle and $\hbar$ the Plank's constant, equation (14) becomes the Schrodinger equation [10]

$$i\hbar \frac{\partial u}{\partial t} + \frac{\hbar^2}{2m} \nabla^2 u = f(x,t) \qquad (18)$$

to define a particle density. $|u(x,t)|^2 \Delta u$ denotes the probability of the particle being in the neighborhood $u(x,t)$ [10]. The fundamental solution of the Schrodinger equation is [5]

$$u_{S_n}^*(x-\xi,t-\tau) = -H(t-\tau)\frac{1+i}{\hbar\sqrt{2}}\left[\frac{m}{2\pi\hbar(t-\tau)}\right]^{n/2} e^{im\|x-\xi\|^2/2\hbar(t-\tau)}. \qquad (19)$$

We have the corresponding DFW transforms

$$S(n,\xi,\tau) = \int_0^{+\infty} \int_{IR^n} V(x,t) \frac{u_{S_n}^*(\xi-x,\tau-t)^{-1}}{\|\xi-x\|} dxdt \qquad (20a)$$

and

$$V(x,t) = \frac{1}{C_S} \int_0^{\infty} \int_0^t \int_{IR^n} S(n,\xi,\tau) u_{S_n}^*(x-\xi,t-\tau) n d\xi d\tau dn. \qquad (20b)$$

Dimension $n$ in (20) also can be the complex number. Now we go to the hyperbolic wave equation

$$\frac{\partial^2 u}{\partial t^2} - c^2 \nabla^2 u = f(x,t), \qquad (21)$$

where $c$ is the media wave velocity. The wave potential with density $f$ is defined by [5]

$$W_n = u_{w_n}^* * f, \qquad (22)$$

where * denotes the convolution operation and $u_{w_n}^*$ is the time-harmonic fundamental solution of wave equation (21). In terms of (22), we can create the corresponding wave potential DFW transforms in terms of dimensionality in the same way as did with the heat potential.

We can also make the DFW series in terms of discrete dimension parameters. However, the basis functions should not be singular, e.g. nonsingular general solution of the Helmholtz equation. It is noted that the higher-dimension basis functions tend to be more compactly supported. Thus, the corresponding interpolation matrices tend to be sparser in the higher dimensions which offsets the so-called the curse of dimensionality.

## 3. Multiple reciprocity DFW transforms

This section is concerned with the DFW transforms using the order of PDE as a substitute of the scale parameter. Since this kind of the DFW is essentially related to the boundary particle method [3,4], a boundary-only numerical technique based on the multiple reciprocity principle [11] to solve a PDE problem, we call it the multiple reciprocity (MR) DFW. To clearly illustrate our idea, consider the following example without a loss of generality

$$\Re\{u\} = f(x), \qquad x \in \Omega, \tag{23}$$

$$u(x) = R(x), \qquad x \in S_u, \tag{24}$$

$$\frac{\partial u(x)}{\partial n} = N(x), \qquad x \in S_T, \tag{25}$$

where $\Re$ is a linear differential operator, $x$ means multi-dimensional independent variable, and $n$ is the unit outward normal. The solution can be expressed as

$$u = u_h^0 + u_p^0, \tag{26}$$

where $u_h^0$ and $u_p^0$ are the zero-order homogeneous and particular solutions, respectively. The multiple reciprocity method evaluates the particular solution in Eq. (26) by a sum of higher-order homogeneous solution, namely,

$$u_p^0 = \sum_{m=1}^{\infty} u_h^m, \tag{27}$$

where superscript $m$ is the order index of the homogeneous solution. Thus, the BPM solution of inhomogeneous equation (23) can be expressed as

$$u = u_h^0 + u_p^0 = \sum_{m=0}^{\infty} u_h^m. \tag{28}$$

We do not want to plunge the readers into details of the BPM and recursive multiple reciprocity solution procedure which can be found in [3,4]. The final solution is given by

$$u(x_i) = \sum_{m=0}^{M} \sum_{k=1}^{L} \beta_k^m u_m^\#(x - x_k), \tag{29}$$

where $u_m^\#$ is the general or fundamental solution of operator $\Re^m\{\}$ which denotes the *m*-th order of $\Re\{\}$, say $\Re^0\{\}=\Re\{\}$, $\Re^1\{\}=\Re\{\Re\{\}\}$. For instance, the general and fundamental solutions of the m-order Helmholtz equation $\Re^m\{\ \} = (\nabla^2 + \gamma)^m$ are given by

$$u_{H_n^m}^\#(r) = A_m(\gamma r)^{-n/2+1+m} J_{n/2-1+m}(\gamma r), \tag{30a}$$

$$u_{H_n^m}^*(r) = A_m(\gamma r)^{-n/2+1+m} H_{n/2-1+m}^{(1)}(\gamma r), \tag{30b}$$

where $A_m = A_{m-1}/(2*m*\gamma^2)$, $A_0 = 1$, and *n* is the dimension of the problem; *J* represents the Bessel function of the first kind and $H^{(1)}$ is the first kind of the Hankle function. It is noted that the recursive multiple reciprocity method described in [3,4] can be directly used with the boundary integral formulation in the same way as in the collocation BPM and will greatly improve computing efficiency of the so-called multiple reciprocity BEM.

It is obvious that (29) is a DFW series in terms of integer "scale parameter" *m*. To introduce the continuous DFW transforms, we need to use the concept of the fractional order of derivative [12], i.e., *m* could be a real number rather than an integer. Thus, we could create the multiple reciprocity DFW transforms with the high-order fundamental solutions of the operator $\Re\{\}$

$$M_n(m,\xi) = \int_{IR^n} f(x)\overline{u^*_{H_n^m}(\xi - x)}dx \tag{31a}$$

and

$$f(x) = \frac{1}{C_M}\int_0^{+\infty}\int_{IR^n} M_n(m,\xi)u^*_{H_n^m}(x - \xi)md\xi dm, \tag{31b}$$

The multiple reciprocity DFW transforms (31) reveal the distribution of function $f(x)$ with respect to the order of the Helmholtz operator. Observing the MR DFW series (29), the MR DFW can also alternatively be given by

$$\Pi_n(m,\xi) = \int_{IR^n}\left\{(\nabla^2 + \lambda^2)^m f(x)\right\}\overline{u^*_{H_n^0}(\xi - x)}dx \tag{32a}$$

and

$$f(x) = \frac{1}{C_\Pi}\int_0^{+\infty}\int_{IR^n}\Pi_n(m,\xi)u^*_{H_n^0}(x - \xi)md\xi dm, \tag{32b}$$

where $u^*_{H_n^0}$ denotes the fundamental solution of the zero-order Helmholtz $\nabla^2 + \gamma^2$. (32) may be better in some practical uses than (31).

[11,13,14] give the high-order general and fundamental solutions of the Laplace, modified Helmholtz, convection-diffusion, Berger and Winkler equations. Consider the scale-invariant Laplacian, we have its 2D and 3D high-order fundamental solutions [11]:

$$u^*_{L_2^m}(r) = \frac{1}{2\pi}r^{2m}(A_m \ln r - B_m), \tag{33}$$

where $A_{m+1} = A_m/4(m+1)^2$ and $B_{m+1} = (B_m/(m+1) + B_m)/4(m+1)^2$ ;

$$u^*_{L_3^m}(r) = \frac{1}{4\pi(2m)!}r^{2m-1}, \tag{34}$$

where $m$ is the Laplacian order. In section 2, we have constructed the potential Laplacian DFW in terms of dimensionality parameter. Here we instead have the Laplacian MR DFW transforms

$$G_n(m,\xi) = \int_{IR^n} f(x) \frac{u^*_{L^m_n}(\|\xi - x\|)^{-1}}{\|\xi - x\|} dx, \quad (35a)$$

and

$$f(x) = \frac{1}{C_G} \int_{c-i\infty}^{c+i\infty} \int_{IR^n} G_n(m,\xi) u^*_{L^m_n}(\|x - \xi\|) m\, d\xi\, dm, \quad (35b)$$

where complex order $m$ corresponds to the complex-order Laplacian. The alternatives of the MR DFW (35) are given by

$$G_n(m,\xi) = \int_{IR^n} \nabla^{2m} f(x) \frac{u^*_{L^0_n}(\|\xi - x\|)^{-1}}{\|\xi - x\|} dx, \quad (36a)$$

and

$$f(x) = \frac{1}{C_L} \int_{c-i\infty}^{c+i\infty} \int_{IR^n} G_n(m,\xi) u^*_{L^0_n}(\|x - \xi\|) m\, d\xi\, dm. \quad (36b)$$

Comparing the MR Laplacian DFW transforms (35) and (36) with the preceding potential Laplacian DFW transforms (5) and (12), one sees that both look very much similar. Zahle [6] pointed out that the spectral dimension of the Laplacian agrees with the Hausdorff dimension underlying fractal through the fractional derivative. In other words, the dimension and the order of the Laplace equation can be reciprocally connected under certain conditions through the Green function of the Reize potential on a fractal. For a detailed discussion see section 5. We also can develop the high-order potential Laplacian DFW transform in terms of dimension parameter, i.e.

$$P^m(n,\xi) = \int_{IR^n} f(x) \frac{u^*_{L^m_n}(\|\xi - x\|)^{-1}}{\|\xi - x\|} dx, \quad (37a)$$

$$f(x) = \frac{1}{C_P} \int_{c-i\infty}^{c+i\infty} \int_{IR^n} P^m(n,\xi) u_{L_n^m}^*(\|x-\xi\|) n d\xi dn . \qquad (37b)$$

In particular, the Mellin transform can be seen as the degenerate form of the one-dimension high-order Laplacian DFW transforms (37).

## 4. Composite multiple reciprocity DFW series

The singularity of the fundamental solutions of multidimensional PDEs does not cause troubles in the creating of various continuous DFW transforms, but makes it seemingly impossible to construct a complete multiple reciprocity DFW series. We have good reason to allot a separate section to deal with this thorny issue. The basic strategy is to use the Green second identity to evaluate the zero-order singular fundamental solution term and leave the other nonsingular high-order fundamental solution terms to be handled in the normal fashion. The title "composite multiple reciprocity" suggests involving more than one PDE operators in some cases, e.g. composite Laplace and Helmholtz operators, and thus is different from the traditional multiple reciprocity [11] where the mere one type of PDEs of different orders is concerned.

**4.1. Laplacian MR DFW series**

To eliminate the edge effect of the Helmholtz-Fourier (HF) series, section 3.3 of [1] developed the HF series of the form:

$$f(x) = f_0(x) + \sum_{j=1}^{\infty} \sum_{k=1}^{\infty} \beta_{jk} \phi_n(\lambda_j \|x - x_k\|), \qquad n \geq 2, \qquad (38)$$

where $f_0(x)$ and the sum term respectively represent the solutions corresponding to nonzero boundary condition and zero boundary condition, and $n$ is the dimensionality and $\lambda$ the eigenvalues; $\phi_n$ denote the HF basis functions based on the nonsingular kernel general solution of the Helmholtz equation. $f_0(x)$ is evaluated by a Laplace equation (degenerate Helmholtz equation with a zero eigenvalue)

$$\nabla^2 f_0(x) = 0. \tag{39}$$

In terms of the Green second identity, we have

$$f_0(x) = \int_{S^{n-1}} \left\{ \frac{\partial f(x_j)}{\partial n} u^*_{L_n^0}(\|x - x_j\|) - f(x_j) \frac{\partial u^*_{L_n^0}(\|x - x_j\|)}{\partial n} \right\} dx_j, \tag{40}$$

where $u^*_{L_n^0}$ is the fundamental solution of the zero-order Laplacian $\nabla^2 u$, and $S^{n-1}$ is the surface of finite domains. (40) can be easily evaluated by the boundary element method (BEM) with boundary $f(x)$ data, and then $f_0(x)$ at any inner locations can be calculated via (40) again. It is well known that the boundary is the dominant features in various data sets. The Green second identity should be the right tool to capture it. The boundary knot method (BKM) [3,13] is an alternative to the BEM for this task.

[56] discusses the essential concept of the complete fundamental solution. For instance, the 2D Laplacian has the essential fundamental solution $\frac{-1}{2\pi}\ln(r)$ and the complete fundamental solution $\frac{-1}{2\pi}(\ln(r) + C)$, where $C$ is a constant. The standard BEM only uses the former. For the DFW series, this may lead to the missing of the constant term and incomplete basis functions. Thus, in the following DFW series, we may need to add a constant term whenever the completeness concern arises, or instead use the complete fundamental solution in (40). On the other hand, the complete fundamental solution also provides an alternative explanation of the multiquadratic (MQ) type radial basis

functions. For instance, the shifted TPS $\ln(r_k^2 + c_k^2)$ has something to do with the 2D Laplacian complete fundamental solution $\frac{-1}{2\pi}(\ln(r)+C)$, while the MQ $\sqrt{r_k^2 + c_k^2}$ and inverse MQ $1/\sqrt{r_k^2 + c_k^2}$ may respectively underlie the 1D and 3D Laplacian complete fundamental solutions $\frac{1}{2}r+C$ and $\frac{1}{4\pi}\left(\frac{1}{r}+C\right)$.

The above HF series provides an insight to construct the multiple reciprocity Laplacian DFW series

$$f(x) = f_0(x) + \sum_{m=1}^{\infty}\sum_{k=1}^{\infty} \alpha_{mk} u^*_{L_n^m}(\|x - x_k\|), n \geq 2, \tag{41}$$

where $f_0(x)$ is evaluated in the same way as (40), and $u^*_{L_n^m}$ is the fundamental solution of the $m$-th order Laplacian $\nabla^{2(m+1)}u$. Observing (33) and (34) of the 2D and 3D Laplacian high-order fundamental solutions, it is found that when the order $m \geq 1$, these kernel solutions are no longer singular at the origin. Thus, (41) kills the singularity culprit and is called the multiple reciprocity Laplacian DFW series. Note that except of evaluating $f_0(x)$ with the boundary elements or BKM, the nodes could be anywhere inside domain and on the boundary. (41) can be understood splitting $f(x)$ into two parts of $f_0(x)$ and the remainder, where the latter is approximated by the high-order Laplacians.

The MR Laplacian DFW series could be an efficient tool to deal with the aperiodic signals which do not exhibit the periodic behavior. It is also very interesting to note that the order $m$ of the Laplacian plays an analogous role of the polynomial order in the standard polynomial interpolation. In fact, expansion series (41) is a multidimensional analog of the univariate polynomial approximation, where the order of monomials is interpreted as the order of the fundamental solution of the one-dimension high-order Laplacian. It is worth pointing out that $u^*_{L_n^m}$ of different orders are orthogonal, which greatly reduces computing effort. The Gram-Schmidt orthogonality method may be helpful to enforce the orthogonalization of the Laplacian basis functions of the same order over translates. The translation invariant polynomial DFW series presented in [2] can also be efficiently calculated in the same recursive way as the MR DFW series for data processing and numerical PDEs.

**4.2. Composite multiple reciprocity DFW series with the solutions of composite PDE**

[1] noted that the nonsingular general solution of the convection-diffusion problem was not suitable to handle exterior unbounded domain problems, while its fundamental solution capable for this task could not be used to create the DFW series due to its singularity at the origin. A solution of this perplexity is to construct the composite MR DFW series with the high-order solutions of a composite PDE.

To illustrate the basic strategy clearly, let us start with the equation

$$\nabla^2 \left( D\nabla^2 u + \vec{v} \bullet \nabla u - ku \right) = -\Delta_i, \qquad (42)$$

where $\vec{v}$ denotes velocity vector, $D$ is the diffusivity coefficient, and $k$ represents the reaction coefficient. (42) combines the convection-diffusion and Laplace equations. Its fundamental solutions are

$$u^*_{C_2^0}(r_k) = \frac{-1}{2\pi} \left( \ln(r_k) + e^{-\frac{\vec{v} \cdot (x-x_k)}{2D}} K_0(\rho r_k) \right), \qquad (43a)$$

$$u^*_{C_n^0}(r_k) = \frac{r_k^{2-n}}{(n-2)S_n(1)} + e^{-\frac{\vec{v} \cdot (x-x_k)}{2D}} \frac{(2\pi\rho r_k)^{-(n/2)+1}}{2\pi} K_{(n/2)-1}(\rho r_k), \quad 3 \le n, \quad (43b)$$

where dot denotes the inner product of two vectors;

$$\rho = \left[\left(\frac{|\vec{v}|}{2D}\right)^2 + \frac{k}{D}\right]^{\frac{1}{2}}. \qquad (44)$$

The fundamental solution (43) is singular at the origin. The high-order fundamental solutions of (42) are a sum of those of the convection-diffusion equation and the Laplace equation. It is found [14] that the high-order fundamental solutions of the convection-diffusion equations except in 2D case are also singular in contrast to the nonsingular high-order fundamental solutions of the Laplacian. Due to this observation, we construct a composite multiple reciprocity series for a continuously differential function $Q(x)$

$$Q(x) \cong Q_0(x) + \sum_{m=1}^{M}\sum_{k=1}^{N} c_{mk} u^*_{L_n^m}(\|x - x_k\|), \qquad n \geq 2, \qquad (45)$$

where $Q_0$ satisfies (42) and is evaluated by the Green second identity as in (40). In data processing, the Dirichlet boundary data are often only accessible. We can use the Green second identity twice, namely, first with the Laplacian to get the Neumann data, and then with the fundamental solution (43) for all necessary high-order boundary data. The expansion (45) is in fact equal to the iterative differentiation of function $Q(x)$ by

$$\nabla^{2M}(D\nabla^2 + \vec{v}\bullet\nabla - k)Q = \nabla^{2M} F(x) = q(x), \qquad (46)$$

where $F(x)$ is the inhomogeneous term of the convection-diffusion equation. If the residues $q(x)$ continuously tends to zero with increasing $m$, (45) will converge. It is noted that expansion (45) includes the direction vector $\vec{v}$ and thus is also well suitable to handle the track data. If $Q(x)$ is due to a quasi-periodic source $P(x)$ consisting of a finite number of periodic components [15], a composite operator of the convection-diffusion, Laplace, and Helmholtz equations of the form

$$\nabla^{2M}(\nabla^2 + \lambda_1^2)\ldots(\nabla^2 + \lambda_S^2)(D\nabla^2 + \vec{v}\bullet\nabla - k)Q = \nabla^{2M}(\nabla^2 + \lambda_1^2)\ldots(\nabla^2 + \lambda_S^2)P(x) = g(x) \quad (47)$$

will be preferred, where $g(x)$ should be smooth and close to a constant or zero. The corresponding DFW series is

$$Q(x) \cong Q_0(x) + \sum_{m=1}^{M}\sum_{k=1}^{N} \alpha_{mk} u^*_{L_n^m}(\|x-x_k\|) + \sum_{i=1}^{S}\sum_{k=1}^{N} \beta_{ik} u^{\#}_{H_n^0}(\lambda_i \|x-x_k\|), \quad n \geq 2, \quad (48)$$

where $u^{\#}_{H_n^0}$ is the general solution of the zero-order Helmholtz equation. The strategy can be extended to handle other problems via a varied combining of different PDEs. We may call it the differentiation smoothing. Following are some composite PDE examples of maybe highly interest:

$$\nabla^{2M}\left(\nabla^2 \pm \mu^2\right)u, \tag{49a}$$

$$\nabla^{2M}\left(\nabla^4 \pm \kappa^2\right)u, \tag{49b}$$

$$\frac{\partial^N}{\partial t^N}\left(\frac{1}{\eta^2}\frac{\partial}{\partial t} - \nabla^2\right)u, \tag{49c}$$

$$\frac{\partial^N}{\partial t^N}\left(\frac{1}{c^2}\frac{\partial^2}{\partial t^2} - \nabla^2\right)u. \tag{49d}$$

(49a) under $m=1$ is the Burger equation governing large deflections of plate without planar displacements which combines the Laplacian and the modified Helmholtz equation. [13] gives some composite PDEs and their high-order fundamental and general solutions. (49c,d) are the temporal multiple reciprocity method which aims to eliminate the inhomogeneous terms due to time-dependent sources. We have a great deal of freedom to combine any PDEs such as the convection-diffusion, Helmholtz, modified Helmholtz, Laplace, elastostatics, time-dependent diffusion and wave and transport equations to tackle a very broad variety of type data processing and PDE problems. The corresponding fundamental solutions will be simply a sum of all individual fundamental solutions. The composite multiple reciprocity approach is also very useful in the kernel distance function [3] to handle domain integral and particular solution of PDE subjected to different types of inhomogeneous function terms.

**4.3. Time-space MR DFW series**

The preceding DFW transform and series are only concerned with spatial variables and equilibrium data. It is feasible to develop the time-space MR DFW series and transforms due to the time-harmonic solution of the diffusion and wave equations, which behaves similarly as the spatial Laplacian harmonic functions. For brevity, we simply display an analogous time-space MR DFW series for wave problems below

$$f(x) = f_0(x,t) + \sum_{m=1}^{\infty}\sum_{k=1}^{\infty} \alpha_{mk} u^*_{L_n^m}\left(\sqrt{c^2(t-t_k)^2 - r_k^2}\right) H\left(c^2(t-t_k)^2 - r_k^2\right), n \geq 2, \quad (50)$$

where $c$ is the wave velocity, $H$ is the Heaviside step function augmented for the causality, and $r_k = \|x - x_k\|$. $f_0(x,t)$ can be evaluated by the time-space boundary element as in (40).

**4.4. Boundary-only MR DFW for data processing and numerical PDE**

Inverse problems are much often common in engineering computation than direct problems. The boundary element method is found advantageous over the domain-type numerical techniques such as the finite elements and finite differences in handling inverse problems since the boundary data in most cases are dominant in determining the systematic behavior and much more easily accessible than the inside-domain data. However, we unfortunately find that the BEM is seldom used in practice to handle inverse problem and data processing problems. This may be due to a few factors: 1) singular integration, 2) full and asymmetric interpolation matrix, 3) costly domain integral, 4) mathematical difficulty. The preceding composite MR DFW series seems a promise to remedy all these defects of the BEM and could be very efficient if the proper composite PDE is found. As well, the boundary particle method may be more promising to handle data processing by only using the boundary data. [3,4,14] shows that the BPM could be symmetric and nonsingular and truly boundary-only for inhomogeneous PDE problems. The wavelet threshold shooting of the composite MR DFW series will lead to a sparse interpolation matrix. On the other hand, the advantages of the MR-BEM over the BPM are twofold: 1) the integral formulation of the MR-BEM is generally more stable than the collocation BPM; 2) the MR-BEM can use the singular fundamental solution without requiring the fictitious boundary outside physical domain. Thus, the BPM and BEM have their respective relative strengths and weaknesses. In conclusion, the BEM and BPM with the help of the recursive multiple reciprocity DFW series are two boundary-only techniques not only for PDE problems but also for data processing.

The basic procedure in the multiple reciprocity solution of inhomogeneous PDE is a reversely recursive iteration [3,4]. Different from the solution of PDE problems, the data processing, however, is a directly recursive iteration, namely, we start to interpolate the accessible boundary data with the zero-order fundamental solution or general solution, and attain the derivative boundary data of one order higher, and then interpolate these obtained data with high-order fundamental or general solution and evaluate the higher-order derivative boundary data, and repeat this processing upward until the residues norm is under acceptable magnitude. For details on the BPM and recursive multiple reciprocity method see [3,4,14]. In this way, we can effectively interpolate the whole domain data only by the boundary data. If the composite multiple reciprocity method presented in section 4.2 is used, we need to evaluate a finite number of interpolation matrices corresponding to each different element PDE. In addition, [57] presents a generalized boundary element method to transfer the domain integral caused by the inhomogeneous media to the boundary integral. The basic strategy behind this method can be applied with the DFW in handling the inhomogeneous media problems.

One may argue that the MR DFW series using the solutions of high-order PDE may not be suitable for discontinuous problems, e.g. functions $F(x)$ in (46) and $Q(x)$ in (47) exhibit discontinuous or low continuous property. The author thinks that this discontinuity issue could be solved by the subregion approach. In cases of concentrated loading, the distance function can be used in the same very simple fashion as in the BEM [16]. It is also stressed that the MR DFW as well as other DFW's are not necessarily global approximation in correspondence to local and global boundary integral equations.

## 5. Connections among dimension, scale and PDE order

In many scientific disciplines, the dimension, scale (frequency) and PDE order are treated as independent parameters. Observing the dimension, PDE order and scale DFW's developed in the present report series, one may wonder if these three basic parameters exhibit inherent connections just like what relationships between mass and energy, time and space are. Recent years witness intense research on applications of fractional derivatives [12], fractional Fourier transform [17] and fractal to many real world problems. Now it is known that there is a definite connection between fractional Fourier and fractional derivative, namely,

$$FT_+\left(\frac{\partial^s p(t)}{\partial t^s}\right) = (-i\omega)^s P(\omega), \qquad (51)$$

where $FT_+$ indicates the Fourier transform, $\omega$ represents frequency, and $s$ is the derivative order, not necessarily an integer, and $P(\omega)$ the Fourier transform of $p(t)$. In the following we will establish the links between fractional derivative and fractal through the analysis of the acoustic frequency dependent attenuation. The power law of attenuation [18] is described by

$$E = E_0 e^{-\alpha(\omega)\Lambda x}, \qquad (52)$$

$$\alpha(\omega) = \alpha_0 \omega^y, \qquad y \in [0,2], \tag{53}$$

where $E$ denotes signal energy, $\alpha_0$ and $y$ are media-dependent parameters obtained through a fitting of measured data. Note that $y$ is real. (53) can be restated as

$$y = \frac{\ln \alpha(\omega)/\alpha_0}{\ln \omega}. \tag{54}$$

(54) can be seen as a self-similar object, i.e. scale invariance [19]. Self-similar fractals with parameters $N$ and $n$ are normally described by a power law

$$N = q^n, \tag{55}$$

where

$$n = \frac{\ln N}{\ln q} \tag{56}$$

is known as the Hausdorff dimension. Note that $q$ here could be any physical parameters besides time and space. Comparing (54) with (56), it is clear that $y$ represents fractal dimensionality in terms of frequency. In other words, the power law attenuation (54) on different frequencies underlies the invariant parameter $y$. The author is very curious why fractal is between 0 and 2.

[20] presented the fractional temporal derivative PDE model to describe the power law (53) of frequency dependent attenuation:

$$\nabla^2 p = \frac{1}{c^2} \frac{\partial^2 p}{\partial t^2} + \frac{2\alpha_0}{c^{1+2y}} \frac{\partial}{\partial t} \left| \frac{\partial^y p}{\partial t^y} \right|, \tag{57}$$

where $p$ denotes acoustic pressure and $c$ is acoustic speed. When $y=0$, (57) is reduced to the standard damped wave equation. In 1D case, $y=2$ leads (57) to an acoustic equation very similar to equation (13) in [21]. Here we find fractal $y$ is mirrored by the $y$-order temporal derivative. Namely, (57) suggests that $y$ represents temporal fractal. However, it is observed from (54) that $y$ is independent of frequency (time scale) $\omega$. It is therefore much reasonable to think that $y$ may in fact mean the spatial fractal. For example, $y$ varies with different human body tissues, which have different spatial microstructures. [20] also proposed a spatial fractional partial derivative model:

$$\nabla^2 p = \frac{1}{c^2}\frac{\partial^2 p}{\partial t^2} + \frac{2\alpha_0}{c^{1+y}}\frac{\partial}{\partial t}\left|\nabla^y \bullet p\right|. \tag{58}$$

When $y=2$, (58) is the known augmented wave equation. It is not difficult to derive the fundamental solution of the above fractional wave equation via the Fourier transform. The absolute value of the fractional derivative in (58) can be calculated via the positive definite Laplacian, i.e.

$$\left|\nabla^y \bullet p\right| = \left(-\nabla^2 p\right)^{y/2}. \tag{59}$$

Both attenuation temporal model (57) and spatial model (58) involving fractional order derivatives show that the damping behavior has much to do with the fractal structure of media, while the differential orders of inertia and diffusion terms are independent of fractional dimensional effect, no matter either integer or fractal dimensions. It is clearly recognized from the above analysis that the dimensional effect on some physical mechanism, e.g. frequency-dependent attenuation, can be exactly reflected by the corresponding derivative order in the PDE model. Now we can conclude that the dimension, scale (frequency) and derivative order are inherently related to describe the attenuation mechanism.

To more clearly clarify the connection between the dimension and derivative order, let us consider the Riesz potential [6,7]

$$I_s f(x) = \frac{\Gamma((n-s)/2)}{\pi^{s/2} 2^s \Gamma(s/2)} \int_{IR^n} \frac{f(\xi)}{\|x-\xi\|^{n-s}} d\xi, \qquad (60)$$

where $s$ is the order of Riesz potential, and $n$ is the dimension. By analogy with the Riemann-Liouville definitions of the fractional derivative [22], we define the fractional Laplacian

$$L^n_{s-y}\{f(x)\} = L^{n+y}_s\{f(x)\} = \frac{\Gamma((n-s)/2)}{\pi^{s/2} 2^s \Gamma(s/2)} \nabla^2 \left\{ \int_{IR^n} \frac{f(\xi)}{\|x-\xi\|^{n-s+y}} d\xi \right\}, \qquad 0 \prec y \prec 2. \qquad (61)$$

(61) clearly shows that the dimension and the order of the Laplacian are inherently connected. By using the multiple reciprocity expansion series, we can have the expansion definition of (62) just like Caputo's expansion definition of the fraction derivative. Two intuitive definitions of the fractional distance variable are given by

$$r_k^s = \left( \sum_{i=1}^n (x^i - x_k^i)^s \right)^{1/s}, \qquad (62a)$$

and

$$r_k^{s-y} = \left( \sum_{i=1}^n (x^i - x_k^i)^{s-y} \right)^{1/(s-y)}, \qquad (62b)$$

where $n$ is the topological dimensionality, and s could even be a complex number. It is noted that $s=2$ leads to the Newtonian potential and Euclidean distance.

Varying $y$ means the multifractal, i.e. continuous or quantum dimension variation, which is visible in some physics problems such as acoustic energy absorption variations over different human body tissues, while a broadband excitation will cause multiscale frequency components. Therefore, a complete description of these complex physical phenomena should involve multifractal, multiscale and multivariate (3M) systems. It is also interesting to see from the attenuation power law (53) that the time scale (frequency) and space dimension are reflected on media acoustic attenuation as what the general relativity reads dependence of time and space on mass.

[20] also noted that the algebra counterpart of the fractional derivative and fractal should be the fractional order of a matrix. For instance, the finite element (FE) numerical discretization (59) is given by

$$\left(-\nabla^2 p\right)^{y/2} = A^{y/2} \bar{p}, \tag{63}$$

where $A$ is the positive definite FE interpolation matrix of the Laplacian, and $\bar{p}$ is the discrete value vector of $p$. From this point of view, the fractional Haar transforms should be the fractional order of the original transform matrices, where the fraction may indicate the fractal dimension or the fractional order derivative. Similarly, we can get the fractional Welsh and Hadamard transforms corresponding to a fractal. Now we have a complete theory of fractional calculus, fractal geometry, and fractional algebra in place to model and analyze complex mathematical physics problems.

Now we try to use the above insights to analyze the electrical current in a transmission line governed by the telegraph equation [10]

$$\nabla^2 u = C \frac{\partial^2 u}{\partial t^2} + (CR + GL) \frac{\partial u}{\partial t} + GRu, \tag{64}$$

where $C$, $G$, $R$, and $L$ are the capacitance, leakage resistance, resistance, and self-inductance per unit length, respectively. These different resistances in the electric current play the same role as does the viscoelastic damping in acoustic wave propagation and mechanical vibration. By analogy with the previous acoustic attenuation PDE model (58), (64) may be reformulated as

$$\nabla^2 u = C\frac{\partial^2 u}{\partial t^2} + \gamma\frac{\partial}{\partial t}\left|\nabla^\tau u\right| + GRu, \tag{65}$$

where $\tau$ is a positive real number to describe the resistance behavior of media and $\gamma$ the alternative collective parameter of *C*, *G*, *R*, and *L*. (65) is different from the fractional telegraph equation given in [23] in that the spatial fractional derivative is used instead of the temporal fractional derivative. It is interesting to imagine that the electric current resistance may obey a power law similar to (52) and (53), where a temperature-related parameter may act like the frequency in damped acoustics.

Furthermore, [24] presented the new complex-order derivative model for frequency dependent storage and damping loss mechanisms more accurately. For example, the following force-displacement model was given for the viscoelastic vibration [24]

$$q(t) + \lambda\frac{d^r q(t)}{dt^r} = v_0\frac{dq(t)}{dt}, \tag{66}$$

where parameters $\lambda$, *r* and $v_0$ are all complex-valued. The Fourier transform of the complex-order derivative is defined by [24]

$$FT_+\left[\frac{d^r q(t)}{dt^r}\right] = (i\omega)^r FT_+[q(t)]. \tag{67}$$

(67) is valid for Re(*r*)>0 and for *n* times differentiable function *q*(*t*) where *n*-Re(*r*)>0. According the preceding analysis, the complex-order *r* may mean the complex dimension. This naturally raises a question about the definition of the distance variable under complex dimension. In research of wave equation

$$\frac{1}{c^2}u_{tt} = u_{xx}, \tag{68}$$

[25] introduced the pseudo-Euclidean distance in terms of Minkowisk 4-dimension space as opposed to the Euclidean distance. Let

$$s = ict, \qquad (69)$$

we have the Laplacian-alike wave potential

$$u_{xx} + u_{ss} = 0. \qquad (70)$$

Its Euclidean distance variable is

$$\hat{r}_k = \sqrt{(x-x_k)^2 + (s-s_k)^2} = \sqrt{(x-x_k)^2 - c^2(t-t_k)^2}. \qquad (71)$$

The author wonders whether the fundamental solution of the complex-order PDE can be expressed in terms of the pseudo-Euclidean distance.

## 6. Shape parameter DFW's

The MQ $\sqrt{r_k^2 + c_k^2}$ and inverse MQ [26], Gaussian $e^{r_k^2/c_k^2}$, the thin plate spline (TPS) $r_k^{2m+1}$ and $r_k^{2m} \ln r_k$ [27] as well as the shifted TPS $r_k^{2m} \ln(r_k^2 + c_k^2)$ [28], a MQ-type TPS, are among the most popular distance functions (radial basis function, RBF), where $c_k$ is called the shape parameter, a pseudo-scale parameter. In general, replacing the Euclidean distance variable $r_k$ by the MQ in the rotational invariant fundamental solution and general solution of PDEs will yield a variety of the MQ-type kernel distance functions [3]. Enormous numerical experiments [29,30] show that the MQ-type distance functions can achieve spectral accuracy if the shape parameter $c_k$ are optimized. Despite intense research has been devoted to analyzing and determining of the shape parameter, unfortunately, $c_k$ is found to be problem dependent and there are not general approaches optimally determining it a priori. But nevertheless [31] proves that the MQ and TPS are conditional positive definite function and [32] shows the MQ could be seen as the pre-wavelets.

On the other hand, Hon and Wu [33] applied the translate-invariant 2D Laplacian harmonic function $e^{-\alpha x^2+\alpha y^2}\cos 2\alpha xy$ as the basis function to construct very simple and efficient numerical scheme for solving the 2D Laplace problems. Just like the MQ, their Laplacian harmonic function also requires determining a problem-dependent parameter $\alpha$ to get the optimal accuracy and convergence speed. In this study, $\alpha$ is also interpreted as the shape parameter. Now we have rotational and translation invariants Laplacian distance functions, both of which include a shape parameter. The purpose of this section is to create these two-type shape parameter distance function wavelets.

**6.1. Rotational invariant shape parameter DFW**

Beside the known MQ, the Possion kernel [34] is also the MQ-type distance function:

$$P_s^n(\|x-x_k\|,s) = \frac{c_n s}{\left(\|x-x_k\|^2+s^2\right)^{(n+1)/2}}, \quad s>0, \tag{72}$$

where $n$ is dimensionality, and $x \in R^n$. $\int P_s^n(x)dx = 1$ for all $s>0$, which mimics the scale function of the multiresolution analysis (MRA). It is, however, stressed that the DFW is not a MRA. Given a function $f \in L^p(R^n)$ ($p \geq 1$), the solution of the Dirichlet problem with boundary value $f$ is [35]

$$u(x,s) = (P_s^n * f)(x) = \frac{2s}{\omega_{n+1}}\int_{IR^n}\frac{f(\xi)}{\left(\|\xi-x\|^2+s^2\right)^{(n+1)/2}}d\xi. \tag{73}$$

It is worth mentioning that since $s$ and $c_k$ play the analogous role under the same potential theory backdrops, the Possion kernel gives an explicit physical explanation of the shape parameter of the MQ. It is also obvious

$$u(x,0) = f(x). \tag{74}$$

A basic result of Littlewood-Paley theory [34] is that

$$\|f\|_p \approx \|g_1(f)\|_p, \qquad 1<p<\infty, \tag{75}$$

where

$$g_1(f)(x) = \left(\int_0^\infty s \left|\frac{\partial u(x,s)}{\partial s}\right|^2 \frac{ds}{s}\right)^{1/2}. \tag{76}$$

Based on the above observations, now we construct the Possion DFW transform

$$H(\xi,s) = \int_{IR^n} \nabla^2 P_s^n(\xi - x, s) f(x) dx. \tag{77}$$

Parameter *s* is understood the scale parameter in the sense of the pre-wavelets. Unlike the MQ, the Possion kernel function (72) has a rapid decay at infinity and is unconditional positive definite function. As such, the MQ itself can also be employed to construct the DFW. Very interestingly, [36] uses the complex shape parameter to stabilize the ill-conditioning MQ interpolation matrix. This reminds us that we could have the complex shape parameter DFW, i.e. *s* in (77) is the complex number.

To get the compactly supported DFW, by analogy with the RBF wavelets given in [37] we introduce the following basis function

$$\psi(\|x-x_k\|,s) = \begin{cases} P_s^n(\|x-x_k\|,s), & \|x-x_k\| \prec s, \\ 0, & \|x-x_k\| \geq s. \end{cases} \tag{78}$$

$\nabla^2 \psi$ could serve the compactly-supported DFW basis function. In addition, $\partial P_s^n / \partial s$ may be also a good DFW basis function. [38] researches the MQ behavior with an infinite shape parameter.

The classic Gaussian RBF can be stated in the following fashion

$$\phi(r_k) = \frac{1}{2\sqrt{\pi\beta}} e^{-r_k^2/4\beta}, \qquad (79)$$

where shape parameter $\beta$ is related to the time translate in the heat potential (16). $\phi(r_k,\beta)$ approaches the Dirac function $\delta(r_k)$ as $\beta \to 0$ [5]. Thus, the shape parameter in (79) is in fact a special form of the preceding time-space heat potential DFW (17). As in the MQ case, $\beta$ can be the complex number.

The diffusion radial basis function presented in [3]

$$\varphi(r_k) = e^{-\alpha r_k}, \qquad (80)$$

is related to the fundamental solution of the 1D diffusion equation. Due to its simplicity, it is very convenient to use (80) creating the shape parameter prewavelet. It is also straightforward to have the corresponding shape parameter DFW series. [39] gave the MQ DFW series. For the Possion kernel DFW, we have the corresponding series

$$f(x) \cong \sum_{j=1}^{N} \sum_{k=1}^{M} A_{jk} P_s^n (\|x - x_k\|, s_j). \qquad (81)$$

**6.2. Translation invariant shape parameter DFW**

Before carrying on, we list some lemmas [33,35] below:

**Lemma 7.1** The shift of harmonic solution (Laplacian solution) is a harmonic function. Any linear combination of harmonic functions is also a harmonic function.

**Lemma 7.2** The derivatives of any harmonic function are also harmonic functions.

**Corollary 7.1**. A sequence of harmonic functions on $\Omega$ which converges locally uniformly on $\Omega$.. Then, the limit function is again harmonic.

[33] uses the 2D translate invariant harmonic function

$$\varphi(x-x_k, y-y_k) = e^{-\alpha(x-x_k)^2 + \alpha(y-y_k)^2 - 2i\alpha(x-x_k)(y-y_k)} \tag{82}$$

with a shape parameter $\alpha$ for numerical solution of the Laplace equation. In fact, there are numerous such translate invariant harmonic functions, e.g.

$$\psi(x-x_k, y-y_k) = e^{\alpha(x-x_k) + i\alpha(y-y_k)} . \tag{83}$$

For higher-dimension Laplacian, we also have similar harmonic functions. These translate invariant harmonic functions can be used as the basis function to construct the DFW transform and series for the function bounded within a finite domain. The author has not a clear idea of physical background of (82) and (83) and wonders if there are the underlying connections between the shape parameters of the rotational and translation invariant Laplacian basis functions.

[2] introduced the harmonic polynomial DFW series using the translate invariant monomial solutions of the high-order Laplacian. It is worth mentioning that such polynomial DFW is not compactly-supported if we do not enforce some additional condition as did (78) to the Possion kernel.

## 7. DFW transforms versus common integral transforms

Besides the foremost outstanding Fourier and Laplace transforms and series, there are many other theoretically and practically important transforms such as the Hartley, Hilbert, Mellin, Abel, Radon, Riemann-Lioville, Weyl, inverse scattering transforms, and their corresponding expansion series, some of which use the eigenfunctions of PDEs. [1] has developed the Helmholtz-Fourier and Helmholtz-Laplace DFW's as the counterparts of the Fourier and Laplace transforms. Section 2 introduced the DFW correspondences of the Mellin and Kontorovich-Lebedev transforms. This section will try to build the DFW correspondences of a few more normal transforms. Furthermore, the DFW methodology might be extendable to group transforms of irregular high-dimensional domains.

It is noted that both the integral equation and integral transform have the similar expression in terms of the kernel function $w(t,a)$, i.e.

$$g(a) = \int_a^b f(t) w(a-t) dt. \qquad (84)$$

Similarly, the distance function wavelet transforms and Green integral are expressed as

$$w(\lambda, \xi) = \int_{IR^n} f(x) g(\xi - x, \lambda) dx \qquad (85)$$

Thus, it is not difficult to make the DFW transforms in connecting to the traditional integral transforms if we know the kernel solution of partial differential equation behind the latter. As examples, we first intuitively conjecture a DFW transform mimicking Weyl transform [40] without any mathematical analysis:

$$W_n(\xi, \gamma) = \frac{2\Gamma(n/2)}{\sqrt{\pi}\Gamma\left(\frac{n-1}{2}\right)} \int_{IR^n} f(x) \left[\|\xi - x\|^2 - \gamma^2\right]^{(n-3)/2} \|\xi - x\| H\left(\|\xi - x\|^2 - \gamma^2\right) dx, \qquad (86)$$

where $H$ is the Heaviside step function. (86) looks like the wave potential DFW transforms discussed in section 2. In terms of the Hilbert transform, a DFW transform is proposed by

$$H_n(\xi,\eta) = \frac{1}{C_n} \int_{IR^n} \frac{f(x)}{\|\xi-x\|-\eta} dx, \tag{87a}$$

$$f(x) = \frac{1}{C_n} \int_0^\infty \int_{IR^n} \frac{H_n(\xi,\eta)}{\eta - \|x-\xi\|} d\xi d\eta, \tag{87b}$$

By analogy with the cylinder Abel transform [41], the corresponding DFW transform is given by

$$A_n(\xi,\kappa) = \frac{1}{C_A} \int_{IR^n} \frac{f(x)\|\xi-x\|}{\sqrt{\|\xi-x\|^2 - \kappa^2}} H\left(\|\xi-x\|^2 - \kappa^2\right) dx, \tag{88}$$

The basis functions of the DFW correspondence of the Hartley transform are

$$g_n(\lambda r_k) = \frac{\lambda^{n-1/2}}{4} (2\pi\lambda r_k)^{-(n/2)+1} \left[J_{(n/2)-1}(\lambda r_k) + Y_{(n/2)-1}(\lambda r_k)\right], \quad n \geq 2, \tag{89}$$

where $Y$ is the Bessel function of the second kind. It is straightforward to construct the DFW transforms based on (89).

In terms of the Stieltjes Transforms, we have the DFW transform

$$S(\xi,t) = \int_{IR^n} \Gamma(p)(\|\xi-x\|+t)^{-p} dx. \tag{90}$$

We need to mention again that the present study is more intuitive conjecture than the rigorous mathematical inference.

In the radial function wavelets (RFW) [40,42,43], the generalized translation operator $T_x$ ($x \geq 0$) for smooth function $\phi(x)$ on $(0,+\infty]$ is defined by

$$T_x\phi(y) = \frac{\Gamma(n/2)}{\sqrt{\pi}\Gamma(n/2-1/2)} \int_0^\infty \phi\left(\sqrt{x^2+y^2+2xy\cos\theta}\right)(\sin\theta)^{n-2} d\theta, \tag{91}$$

where $x$ and $y$ represents two different nodes coordinates. If the above translation operator is redefined by

$$T_x\phi(y) = \phi(\|x-y\|), \tag{92}$$

and then, all the related results of the RFW [40,42,43] can be used with the distance function wavelets.

Further, in the sense of translation invariant ($\phi(x)$ on $(-\infty,+\infty)$) [44]

$$T_x\phi(y) = \phi(x-y), \tag{93}$$

we may get more general results. In addition, $\theta$ in the translation operator (91) may be expressed via both the inner product distance variable and the Euclidean distance variable, namely,

$$\theta = \chi(\|x-y\|, \|x\|, \|y\|, x \cdot y), \tag{94}$$

$$2x \cdot y = \|x\|^2 + \|y\|^2 - \|x-y\|^2, \tag{95}$$

where dot denotes the inner product of two vectors for the ridge distance variable. The other radial transforms and series [45] may be feasible to be transferred into the DFW transforms and series.

## 8. Geodesic DFW

The processing of geodesic data is one of the important applications of the radial basis function [46]. It is, however, noted that there is no particular kernel distance function available now in literature to deal with the geodesic problems. This section will give a few geodesic kernel distance functions based on the fundamental solutions of the related PDEs.

For anisotropic and inhomogeneous objects, [5] lists the fundamental solutions of the n-dimensional time-dependent diffusion equation

$$u^*_{h_n}(x-\xi, t-\tau) = \frac{H(t-\tau)\kappa^{-1/2}}{[4\pi(t-\tau)]^{n/2}} e^{-R^2/4(t-\tau)}, \tag{96}$$

and of the n-dimensional Laplace equation

$$u^*_{L_n}(x-\xi) = \begin{cases} -\dfrac{\kappa^{-1/2}}{S_2} \ln R, & n=2, \\ -\dfrac{\kappa^{-1/2} R^{2-n}}{(n-2)S_n}, & n \neq 2, \end{cases} \tag{97}$$

where the coefficient matrix $\kappa=[\kappa_{ij}]$ represents the parameters in different directions ($j$) and locations ($i$) of anisotropic and inhomogeneous media, and the geodesic distance $R$ is defined by

$$R^2 = \sum_{i,j=1}^n \kappa^{-1}_{ij}(x_i-\xi_i)(x_j-\xi_j). \tag{98}$$

By analogy with the normal Laplacian high-order solutions (33) and (34), it is straight forward to have the high-order fundamental solutions of the anisotropic and inhomogeneous Laplacians. It is also not difficult to get the geodesic fundamental solutions of other PDEs such as the convection-diffusion, Berger plate, Winkler plate, and wave equations. For instance, we can easily derive the fundamental solution of the Helmholtz equation

$$\varphi_n(x-\xi) = \frac{i\kappa^{-1/2}}{4}\left(\frac{\lambda}{2\pi R}\right)^{(n/2)-1} H^{(1)}_{(n/2)-1}(\lambda R), \quad n \geq 2. \qquad (99)$$

In addition, [58] presents the general approach to obtain the Laplacian fundamental solutions in the isotropic heterogeneous media. It is straightforward to construct various geodesic (inhomogeneous) DFW transforms and series by the PDE kernel solutions of (96), (97), (99) and those given in [58].

## 9. Kernel distance sigmoidal functions

In artificial neural network and machine learning, each node has a transfer function which consists of an activation function (a summation of inputs) and an output function. The type of activation function depends strongly on its task [47]. Among various possibilities, a variety of distance variables construct the simplest and efficient activation function. Note that we here use the widest definition of the distance function [1], which includes the inner product in the ridgelet function, various distance functions listed in [47], and a combination of both. The readers have seen from these reports I, II and III that the translation invariant distance vector, inner product, Euclidean distance function and their combinations appeared in the foregoing elliptical, hyperbolic, and parabolic distance function basis functions. [47] argues that the symbolic (nominal) variables can not be in any liner order. A color variable example is discussed, where each different color is assigned an identity value and a linear distance thus makes little sense in this type of data. However, if we use a different digital definition of colors, situations will alter. For instance, we can assign 0, 0.5, 1 for blue, green, red. Then, any other colors can be something between 0 and 1. Under this definition of variables, the distance variable does make sense. The basic requirement is that if the distance measure between two values is small, both should belong to similar classifications [48].

As of the output function, the S-shaped sigmoidal functions may be the most popular [49,50]. The sigmoidal function yields a value between 0 and 1 in terms of the activation function value $A$ and a slope (scale) parameter $s$. The logistic function [49]

$$\sigma(s) = \frac{1}{1+e^{-As}}. \tag{100}$$

illustrates the typical form of the sigmoidal functions.

This section will focus on developing of the kernel sigmoidal functions with the help of the kernel distance functions. It is stressed that although the kernel distance function has the origin of PDEs, any activation function can be used with them to create a transfer function. The singularity at the origin of some kernel distance function is not an issue for the sigmoidal functions.

As an extension of the hyperbolic functions, we first define the multidimensional hyperbolic functions

$$\sin g(x) = \varphi_n^{\#}(r) - \varphi_n^{*}(r), \tag{101a}$$

$$\cos g(x) = \varphi_n^{\#}(r) + \varphi_n^{*}(r), \tag{101b}$$

$$\tan g(x) = \frac{\varphi_n^{\#}(r) - \varphi_n^{*}(r)}{\varphi_n^{\#}(r) + \varphi_n^{*}(r)}, \tag{101c}$$

$$\csc g(x) = \frac{1}{\varphi_n^{\#}(r) - \varphi_n^{*}(r)}, \tag{101d}$$

$$\sec g(x) = \frac{1}{\varphi_n^{\#}(r) + \varphi_n^{*}(r_k)}, \tag{101e}$$

$$\coth g(x) = \frac{\varphi_n^{\#}(r) + \varphi_n^{*}(r)}{\varphi_n^{\#}(r) - \varphi_n^{*}(r)}, \tag{101f}$$

where $\varphi_n$ with superscripts # and * are respectively the general solution and the fundamental solution of the modified Helmholtz equation

$$\varphi_1^{\#}(sr_k) = \frac{1}{2s} e^{sr_k}, \tag{102a}$$

$$\varphi_n^{\#}(sr_k) = \frac{1}{2\pi}(2\pi sr_k)^{-(n/2)+1} I_{(n/2)-1}(sr_k), \qquad n \geq 2, \tag{102b}$$

$$\varphi_1^{*}(sr_k) = \frac{1}{2s} e^{-sr_k}, \tag{103a}$$

$$\varphi_n^{*}(sr_k) = \frac{1}{2\pi}(2\pi sr_k)^{-(n/2)+1} K_{(n/2)-1}(sr_k), \qquad n \geq 2; \tag{103b}$$

where $r_k = \|x - x_k\|$, $n$ is dimensionality, and $I$ denotes the modified Bessel function of the first kind which grows exponentially as $r_k \to \infty$. In contrast, the modified Bessel function of the second kind $K$ decays exponentially and has singularity at the origin. (101) degenerate into the hyperbolic function in 1D case.

The modified Helmholtz sigmoidal function using the fundamental solution is given by

$$\sigma_{MF}(s) = \frac{1}{1 + \varphi_n^{*}(sA)}, \tag{104}$$

where $A$ is the activation function for a summation of inputs. The natural choice of the distance variable is the Euclidean one for the modified Helmholtz equation. This, however, is not necessary. $\sigma_{MF}$ can be simplified as

$$\hat{\sigma}_{MF}(s) = \frac{1}{1 + e^{-sA}/\left[(sA)^{n/2-1} \ln(A)\right]}. \tag{105}$$

The modified Helmholtz sigmoidal function using the general solution is given by

$$\sigma_{MG}(s) = \frac{\varphi_n^\#(sA) - 1}{\varphi_n^\#(sA) + 1}. \tag{106}$$

The modified Helmholtz equation is also called the diffusion equation and exhibits closely relationship with the stochastic dynamics. Thus, the above modified Helmholtz sigmoidal functions are also underpinned by the statistics.

As seen before, the DFW basis functions using the general and fundamental solutions of the convection-diffusion equation naturally combine the inner product for direction vector and Euclidean distance variables, i.e.

$$\phi_n\left(\vec{v}\cdot(x-x_k), \rho\|x-x_k\|\right) = \frac{1}{2\pi} e^{-\frac{\vec{v}\cdot(x-x_k)}{2D}} (2\pi\rho r_k)^{-(n/2)+1} K_{(n/2)-1}(\rho r_k), \tag{107}$$

$$\phi_n^\#\left(\vec{v}\cdot(x-x_k), \rho\|x-x_k\|\right) = \frac{1}{2\pi} e^{-\frac{\vec{v}\cdot(x-x_k)}{2D}} (2\pi\rho r_k)^{-(n/2)+1} I_{(n/2)-1}(\rho r_k). \tag{108}$$

It is straightforward to have the convection-diffusion kernel sigmoidal function of the form

$$\sigma_{CF}(w,s) = \frac{1}{1 + \phi_n^*\left(w^T \cdot x, s\|x-x_k\|\right)}, \tag{109}$$

$$\sigma_{CG}(w,s) = \frac{\phi_n^\#\left(w^T \cdot x, s\|x-x_k\|\right) - 1}{\phi_n^\#\left(w^T \cdot x, s\|x-x_k\|\right) + 1}, \tag{110}$$

where $w$ is the direction vector. The kernel solution $u_{h_n}^*$ (16) of the time-dependent heat equation can be used to create the time-space sigmoidal function

$$\sigma_H(s,\alpha) = \frac{1}{1+u_h^*(s\|x-\xi\|,\alpha(t-\tau))}. \tag{111}$$

In the same way, the fundamental solutions (96) of heterogeneous transient diffusion problem using the geodesic distance variable (98) can be employed in making the time-space sigmoidal functions with anisotropic and inhomogeneous parameters. In addition, the kernel solution of the time-dependent convection-diffusion equation is

$$\phi_n(v\cdot(x-x_k),\rho\|x-x_k\|) = e^{-\frac{\bar{v}\cdot(x-x_k)}{2D}} \frac{H(t-\tau)}{[4\pi\kappa(t-\tau)]^{n/2}} e^{-\|x-x_k\|^2/4\kappa(t-\tau)}. \tag{112}$$

In terms of (112), it is straightforward to construct the convection-diffusion temporal sigmoidal function.

The fundamental solutions of the potential Laplace equations are entitled in creating the cheap sigmoidal functions

$$\sigma(s) = \frac{1}{1+(sA)^{2-n}/(n-2)S_n}, \qquad n\geq 3, \tag{113}$$

which are much more computationally efficient since no costly Bessel functions and exponents are involved. In addition, the fundamental solutions of wave equations are also eligible to make the sigmoidal function. We also can use the fundamental solutions of the Helmholtz equations to construct the periodic sigmoidal functions

$$\sigma_H(s) = \frac{1}{1+H_n^*(sA)}, \tag{114}$$

and

$$\sigma_H(s) = \frac{1-H_n^\#(sA)}{1+H_n^\#(sA)}, \tag{115}$$

where $H_n$ with superscript * and # respectively represent the fundamental and general solutions of the n-dimensional Helmholtz equation. We can also create variants of the above distance sigmoidal functions by analogy with the sigmoidal functions (46-51) in [49]. Among them, the geodesic kernel distance functions (96) and (97) are especially attractive. In addition, most of DFW's and their basis functions in this report series such as the Possion kernel can be directly used as the transfer function in their own right.

It is observed that the S-shaped sigmoidal function behaves very much like what the shock wave does. Therefore, various distance sigmoidal functions presented above can also be used in the numerical solution of shock wave and boundary layer problems. It is very interesting to compare the shock wave, boundary layer and neuron transfer behaviors. For instance, the Peclet number in the convection-diffusion equation, which is the ratio of the bulk heat transfer and conductive heat transfer, plays an analogous role as dose the slope parameter $s$ in the sigmoidal function. The Peclet number is decided by the heat capacity $C_p$, thermal conductivity $k$, density $\mu$, characteristic length $D$ and density $\rho$ of media [51], i.e.

$$Pe = \frac{D \cdot C_p \cdot \mu \cdot \rho}{k}. \tag{116}$$

The author guesses that the electric signals traveling within neuron may also exhibit similar mechanism and obey the similar mathematical equations such as the Navier-Stokes equation or simpler Burgers equation. Thus, the slope parameter $s$ could be decided via a few physical parameters parallel to those in (116) for the Peclet number. On the other hand, the Reynolds number is proportional to the ratio of inertial force and viscous force and is used in momentum, heat, and mass transfer to describe dynamic similarity [52]. It is conceivable that the solutions of the mathematical physics PDE models of brain neuron may depend crucially on some parameters like the Peclet or Reynolds numbers.

## 10. Concluding remarks

A distributed parameter system is represented by a PDE, its counterpart is lumped parameter systems or simply data set. This suggests any discrete signal, no matter whether linear or nonlinear, memoryless and memory, random or deterministic, having or having not PDE model, must have their respective PDE solution structures. So, we should be aware of the lurking PDE backdrop in processing of specific data such as those in social sciences, linguistics, and literature. It is well known that all PDEs can be categorized into elliptical, parabolic, hyperbolic or their mixture types. As such, we must recognize which kind of PDE type the data we are handling belong to. Majority of data approximation techniques available now, e.g. spline, however, have their roots on elliptical Laplace and biharmonic equations and are not suitable and efficient in handling the hyperbolic and parabolic data.

One of the major drawbacks in the standard wavelets is less incapable of including inherently the features of the particular type objects. In contrast, the DFW can be easily designed with the characteristics of certain type problems. For instance, the fundamental solution of Laplacian under an axisymmetric region has the form [53,54]

$$w = \frac{4E(s)}{(p+q)^{1/2}}, \tag{117}$$

where $E$ is the complete elliptic integral of the first kind, and

$$p = x_i^2 + x_k^2 + (y_i - y_k)^2, \tag{118a}$$

$$q = 2x_i x_k, \tag{118b}$$

$$s = \frac{2q}{p+q}, \tag{118c}$$

where $x$ and $y$ are the Cartesian coordinates. With the Laplacian fundamental solution (117), we can easily make the axisymmetric Laplacian DFW transforms and series as in sections 2 and 3.

In dealing with the infinite domain problems, the distance function presented in [54,55]

$$u(r_k) = \frac{2C - r_k}{(r_k + C)^4} \tag{119}$$

satisfies the regularity conditions of diffusion problems at infinity, where $C$ is a problem-dependent constant. It is obvious that (119) and its variants will be nice DFW basis functions for diffusion type problems within infinite domains.

This report as well as the other two [1,2] explored many DFW-related issues intuitively without enough rigorous mathematical justifications. These DFW's share all of the major merits of the standard wavelets. Unlike the latter, the DFW in principle, however, is mathematically very simple and computationally efficient to perform meshfree high-dimensional data processing and numerical PDEs. The essential differences between the DFW and the traditional integral transform and series lie in that the DFW includes the dimensional effect and keeps the translate and/or rotational invariance.

As complements, there are another two reports [59,60] on the kernel distance function and its applications. The readers are advised to read them for a complete comprehension of the distance function. The DFW is expected to find numerous applications in data processing and numerical PDE. For instance, the composite multiple reciprocity and the differentiation smoothing are very effective and efficient techniques in evaluating the domain integral in terms of the BEM and particular solution in terms of the BKM and BPM, while the Helmholtz-Fourier series is a competitive alternative to the Fourier series method for the domain integral. Some important PDEs and their solutions were not mentioned in the three reports. For instance, the kernel solutions of the delay PDEs and the Black-Scholes equation can also be used to construct the important DFW's.

A generalized distance function wavelet transform with multiple arguments is stated as

$$F(\lambda, \bar{w}, n, m, \xi) = \int_{IR^n} f(x) \Re_n(\lambda, \bar{w}, m, \xi - x) dx , \qquad (120)$$

where arguments $\lambda$, $\bar{w}$, $n$, $m$ and $\xi$ represent scale, direction, dimension, order of high-order PDE, and translate, respectively. The wavelet basis function $\Re_n$ could be the kernel distance function solution of a partial differential equation, which best fits the system behavior of interest. The shape parameter discussed in section 7 is considered an expedite substitute of the scale, dimension, or order of PDE and thus is not considered an independent parameter to be included in (120). Multiple parameters DFW transform (120) may be useful to analyze very complex systems, where some or all of those parameters manifest locally.

The ultimate goal of this research is to show some recent advances in building a complete framework of the distance function. Given their immature nature, it is inevitable to have more or less errors in these reports. Some cautions on the given results are necessary based on readers' own discretion to avoid any potential misleading, and the readers may better regard the reports as a gathering of intuitive ideas rather than the established theory.